\documentclass[preprint,12pt]{elsarticle}

\usepackage{graphicx} 
\usepackage{amsmath}
\usepackage{amssymb}

\usepackage{xcolor}
\usepackage{float}
\usepackage{svg}

\usepackage{graphicx}% Include figure files
\usepackage{dcolumn}% Align table columns on decimal point
\usepackage{bm}% bold math
\usepackage[utf8]{inputenc}
\usepackage[T1]{fontenc}
\usepackage{mathptmx}
\usepackage{etoolbox}

\journal{Physica A}

\begin{document}
	\begin{frontmatter}
		\title{Chain Reaction of Ideas: Can Radioactive Decay Predict Technological Innovation?}
		\author{G. S. Y. Giardini}
		\author{C. R. da Cunha}
		\ead{carlo.cunha@nau.edu}
		\affiliation{organization= {School of Informatics, Computing, and Cyber-Systems, Northern Arizona University}, 
			addressline={1295 S. Knoles Dr.},
			city={Flagstaff},
			state={AZ},
			postcode={86011},
			country={USA}			
		}
		 
		\date{\today}
		
		\begin{abstract}
			This work demonstrates the application of a birth-death Markov process, inspired by radioactive decay, to capture the dynamics of innovation processes. Leveraging the Bass diffusion model, we derive a Gompertz-like function explaining the long-term innovation trends. The validity of our model is confirmed using citation data, Google trends, and a recurrent neural network, which also reveals short-term fluctuations. Further analysis through an automaton model suggests these fluctuations can arise from the inherent stochastic nature of the underlying physics. 
		\end{abstract}
		
		\begin{keyword}
			Markov Chain\sep Cellular Automata\sep Neural Networks
		\end{keyword}

\end{frontmatter}

%================================================
% SECTION
%================================================
\section{Introduction}
The radioactive decay of a specified isotope, such as $^{238}_{92}$U undergoing a decay into $^{234}_{90}$Th $+$ $^4_2$He, can be effectively modeled as a stochastic Markov process \cite{nuclear1}. Unstable nuclei spontaneously decay and emit radiation (e.g., $\alpha$-particles in this process) at a rate characterized by the death transition rate $\mu$. Simultaneously, other stable nuclei in the vicinity may capture these emitted particles and become unstable, initiating a new decay chain. This capture process is modeled by the birth transition rate $\lambda$. Both $\lambda$ and $\mu$ depend solely on the current number of remaining nuclei (denoted by $n$) and exhibit the memoryless property characteristic of Markov chains, meaning the probability of a decay event depends only on the current state $n$, not on the history of previous decays.

Building upon the Bass diffusion model \cite{Bass,Bass2,Bass3}, this work presents a novel approach to modeling the innovation process inspired by the dynamics of radioactive decay. While the Bass model identifies innovators (early adopters) and imitators (influenced by existing adopters), we propose a birth-death Markovian framework focusing on the agents who either adopt or abandon a specific technology. Analogous to radioactive decay coefficients, our model utilizes birth and death transition rates governed by the current number of adherents. This framework transcends the Bass model's emphasis on imitation by capturing the inherent stochasticity of adoption and abandonment decisions, enabling a more nuanced understanding of the innovation process dynamics.

Several existing approaches attempt to model innovation dynamics. Some strategies are based on cognitive processes as random walks on complex networks, where nodes symbolize ideas \cite{IdeaNet}. Catastrophe theory has also been employed to understand innovation through the lens of cusp geometry \cite{catas}. Population dynamics, especially those utilizing models influenced by social regularities and information flows, offer another avenue for understanding the growth patterns elucidated in this research paper \cite{PRE}.
Additionally, various models leverage frameworks such as epidemics, probit regressions, information cascades, competition, and population dynamics \cite{InoRev,InoRev2,InoRev3}. 

Our proposed model distinguishes itself from these methods by establishing a clear analogy to radioactive decay processes through the formal framework of Markov chains. This direct connection with well-understood physical phenomena offers unique advantages in terms of interpretability and potential predictive power.

In the next section, we discuss our model in detail and present simple ansatzes for modeling the birth and death, or attachment and detachment rates. We will show that our estimates lead to a Gompertz-like function \cite{Gompertz} that models the number of active species. We test our model in the following section with both the number of scientific papers and cultural trends captured by Google searches. The estimates for the birth and death curves are then tested with a recursive neural network, which also predicts fluctuations in the model. The last section examines the origins of these fluctuations using an automaton to model both the radioactive and the innovation processes.

%================================================
% SECTION
%================================================
\section{The Model}
\label{chap:The_Model}

Our model considers a population of individuals that either attach to or detach from a technology.
The state corresponding to a population of $n$ agents adhering to a new technology is mapped to the stochastic process $X_n$. The attachment rate per single cell for the new technology is given by $\lambda_n(t)$, whereas the detachment rate is given by $\mu_n(t)$. Moreover, due to social conformity \cite{Banerjee1992,EuEcon}, the chance an individual attaches to or detaches from a technology depends only on the number of individuals already attached to it. Therefore, we choose $\lambda_n(t)=n\lambda_0(t)$ and $\mu_{n}(t)=n\mu_0(t).$

We make the assumption that the stochastic process $X_n$ adheres to the Markov property, which asserts its independence from its entire historical trajectory. Consequently, given the information available up to time $n$, denoted by a set of observations $x_n$ and encapsulated within the sigma-algebra $\mathcal{F}_{n}=\sigma(X_1=x_1,X_2=x_2,\hdots,X_{n-1}=x_{n-1})$, the Markov property asserts that the conditional probability of $X_{n+1}$ at time $n+1$ given $\mathcal{F}_n$ is equivalent to the conditional probability of $X_{n+1}$ given the immediate preceding state $X_n$. Mathematically, this is expressed as $P(X_{n+1}=x_{n+1}|\mathcal{F}_{n})=P(X_{n+1}=x_{n+1}|X_n=x_n)$.

The process $X_n$ induces a Markov chain, reflecting its dependence on its current state. This dependence is captured by three components:

\begin{enumerate}
	\item{Persistence:} The initial term, $(1-\lambda_n\Delta-\mu_n\Delta_t)X_n(t)$, captures the likelihood of remaining in the current state ($X_n$). This term encompasses scenarios where there is no radioactive decay/particle transfer, reflecting the absence of individuals either adopting or abandoning the technology. Here, $\lambda_n$ represents the birth rate at instant $n$, and $\mu_n$ denotes the death rate at the same instant.
	\item{Adoption:} The second term, $\lambda_{n-1}(t)\Delta_t X_{n-1}(t)$, describes the increase in state $X_n$ resulting from transitions originating in preceding state $X_{n-1}$. This term analogously reflects the receipt of radioactive particles or the adoption of new technology by individuals, where the influence is quantified by the birth rate associated with the prior state and the duration of the time interval $\Delta_t$.
	\item{Abandonment:} The third term, $\mu_{n+1}(t)\Delta_t X_{n+1}(t)$, represents the reduction in state $X_n$ attributed to transitions towards the subsequent state $X_{n+1}$. Analogous to unstable nuclei emitting particles or individuals disengaging from the technology, this reduction is contingent upon the death rate associated with the succeeding state.
\end{enumerate}
Mathematically, these concepts are combined in the following equation:

\begin{eqnarray}
	X_n(t+\Delta_t)=&&(1-\lambda_n\Delta-\mu_n\Delta_t)X_n(t)+\lambda_{n-1}(t)\Delta_t X_{n-1}(t)\nonumber\\
	&&+\mu_{n+1}(t)\Delta_t X_{n+1}(t).
\end{eqnarray}
At the limit of very short intervals, $\Delta_t\rightarrow 0$, this equation becomes:

\begin{eqnarray}
	\frac{dX_n(t)}{dt}=&&-(\lambda_n(t)+\mu_n(t))X_n(t)+\mu_{n+1}(t)X_{n+1}(t)\nonumber\\
	&&+\lambda_{n-1}(t)X_{n-1}(t).
\end{eqnarray}

While this equation focuses on individual transitions, one important question arising by this model pertains to the average number of individuals engaged with the technology at any given time. This average engagement is captured by the following expression:

\begin{equation*}
	M(t)=\sum_{n=1}^\infty nX_n(t).
\end{equation*}
The rate of change of this average engagement is determined by:

\begin{eqnarray}
	\frac{dM(t)}{dt}&&=-\left(\lambda_0(t)+\mu_0(t)\right)\sum_{n=1}^\infty n^2X_n(t)\nonumber\\
	&&+\mu_0(t)\sum_{n=1}^\infty n(n+1)X_{n+1}(t)+\nonumber\\
	&&+\lambda_0(t)\sum_{n=1}^\infty n(n-1)X_{n-1}(t)\nonumber\\
	&&=\left(\lambda_0(t)-\mu_0(t)\right)M(t).
	\label{eq:Adherence_Variation}
\end{eqnarray}

Building upon the understanding of the average adoption level from the transition probabilities, let's now delve deeper into the temporal dynamics of technology engagement. To capture the evolving landscape of opinions and preferences, we propose that the attachment rate ($\lambda_0$) representing new users joining may decrease over time due to factors like competing technologies, changing trends, and fading novelty.  We suggest modeling this decrease using a simple hyperbolic function given by $\lambda_0(t)=\alpha/t$, where $\alpha$ is a characteristic time constant. This simplified model employs a functional form akin to pulsed neutron-induced activation \cite{neutron2,neutron3,neutron4}, where the decay rate may undergo transient changes over time \cite{neutron1}. While the actual birth rate dynamics in the system can be intricate, this model effectively captures the initial trend and facilitates the derivation of analytical solutions.

Conversely, the detachment rate $\mu_0$ represents users abandoning the technology, influenced by various factors. We expect this rate to follow a diffusion pattern, with rapid initial adoption followed by a gradual slowdown and stabilization at a long-term average detachment rate. This characteristic resembles the behavior of certain radioactive processes, like pulsed neutron-induced activation, where temporary changes in activation rates might occur due to external influences but the underlying dynamics remain governed by a constant decay constant. To capture this mean-reverting behavior, we propose employing an Ornstein-Uhlenbeck (OU) process \cite{EuEcon}, described by $d\mu_0(t)=\frac{\beta-\mu_0(t)}{\sigma}dt+\kappa dW$, where $\sigma$ is the mean-reversion time, $\beta$ is the equilibrium death rate, $\kappa$ is a volatility parameter, and $dW$ is a Wiener process. 

The death rate may initially be low but gradually increases as the initial excitement wanes and limitations become apparent. Given our interest in major trends, we assume $\kappa\sigma<<1$ and investigate the quasi-deterministic regime of the rate. This simplification facilitates the analysis, allowing us to focus on the general trend of the death rate. Furthermore, we assert that the death rate at the beginning of the process is zero, as there has been no adherence to the technology yet. Under these conditions, the OU process predicts a death rate $\mu_0(t)=\beta\left(1-e^{-t/\sigma}\right)$.

Given these considerations about the birth and death rates, the time variation of the expected number of individuals attached to the technology becomes:

\begin{equation}
	\frac{dM(t)}{dt} = \left(\frac{\alpha}{t}-\beta\left[1-exp\left( -\frac{t}{\sigma}\right)\right]\right)M(t).
\end{equation}
The solution of this equation is:

\begin{equation}
	M(t)=M_0 t^\alpha e^{-\beta( \sigma e^{-t/\sigma} + t)}.
	\label{eq:number_of_attachments_over_time}
\end{equation}
It is interesting to note that the exponential component of the equation is a Gompertz function \cite{Gompertz}, that describes, for example, the population dynamics in confined spaces \cite{cspace}.

Parameters $M_0$, $\alpha$, $\beta$, and $\sigma$ can be found using a standard Levenberg - Marquardt fitting \cite{CRC_ML_Book}. These parameters allow us to compare the temporal dynamics of different technologies.

	%....................................................................
	% SUBSECTION
	%....................................................................
	\subsection{Results} 
	\label{chap:Innovation_Characterized_From_Data_And_Model}
	
	To validate our theoretical model's ability to capture real-world dynamics, we analyzed two contrasting datasets: scientific impact measured through citation counts for three physics papers published in $1980$ (A--C \cite{LeGuillou1980,Ambegaokar1980,Klein1980}) and popular cultural trends represented by Google Trends interest data for ``facebook'' and ``snapchat'' spanning 20 years (2004--2023). We retrieved the citation data using the Web of Science citation report tool over a $44$-year period. Google defines interest as the worldwide number of searches relative to the highest point over the time series on a base $100$. To ensure robust comparisons, both datasets were normalized and included only non-zero data points.
	
	%==================================
	% FIGURE
	%==================================
	\begin{figure}[H]
		\centering\includegraphics{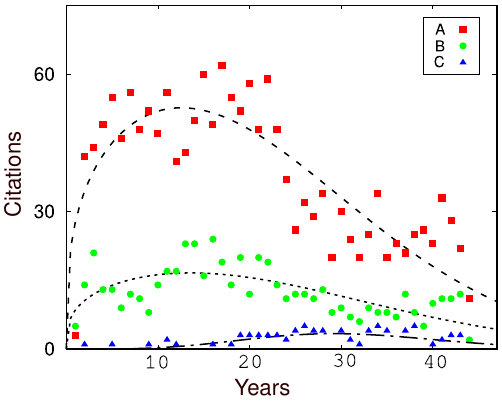}
		\caption{Annual citation counts for three scientific papers (A--C \cite{LeGuillou1980,Ambegaokar1980,Klein1980}) are shown. Square, circle, and triangle symbols denote experimental data for each paper, respectively. Model predictions based on Eq. \ref{eq:number_of_attachments_over_time} are presented as dashed and dotted curves.}
		\label{fig:citations_fit}
	\end{figure}
	
	\begin{table}[H]
		\centering
		\caption{Fitting parameters and the root mean square (RMS) error of the residuals for three scientific papers (A--C \cite{LeGuillou1980,Ambegaokar1980,Klein1980})}
		\begin{tabular}{|c|c|c|c|c|c|}
			\hline
			\# of Citations & RMS of Res. & $\sigma$ & $\beta$ & $\alpha$ & $M_{0}$    \\ \hline
			A & 8.2545 & 1.27918 & 0.0622717 & 0.644351 & 23.1076 \\ \hline
			B & 4.06217 & 17.924 & 0.0632812 & 0.40297 & 23.6886   \\ \hline
			C & 1.33186 & 37.5037 & 0.302128 & 4.67661 & 0.58463   \\ \hline
		\end{tabular}
		\label{tab:first_fit}
	\end{table}
	
	Figure \ref{fig:citations_fit} demonstrates the model's ability to capture diverse citation patterns, revealing both rapid (Paper A) and gradual (Paper C) initial citation growth. These contrasting patterns align with two distinct innovation models: the Socratic and Schumpeterian models  \cite{Lee2022}. The Schumpeterian model is characterized by a slower, more deliberative initial growth in citations, while the Socratic model is marked by a faster, disruptive initial increase. Notably, our model accurately captures the citation dynamics of both innovation types, as evident in the figure and the estimated parameters presented in Table \ref{tab:first_fit}.
	
\begin{figure}[H]
		\centering\includegraphics{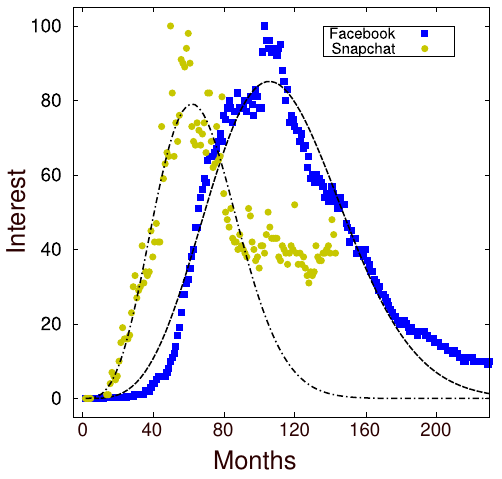}
		\caption{Temporal evolution of interest in the keywords `Facebook' and `Snapchat,' featuring experimental data represented by square and circle symbols. Model predictions using Eq. \ref{eq:number_of_attachments_over_time} are depicted by the dashed curves.}
		\label{fig:social_media}
	\end{figure}

	Figure \ref{fig:social_media} expands our analysis to popular cultural trends, as reflected in Google Trends data for the keywords ``Facebook'' and ``Snapchat''. The corresponding estimated parameters are presented in Table \ref{tab:second_fit}. While the overall fitting results are satisfactory, noteworthy divergences appear after month $100$ for ``snapchat'' and after month $190$ for ``Facebook''. Remarkably, these points in time coincide with the latter half of $2019$, aligning with the onset of the impact of COVID-$19$. This divergence suggests that the pandemic can be considered a perturbation, disrupting the natural trajectory of innovation adoption and detachment. In the context of our radioactive analogy, this scenario resembles the introduction of a new neutron-induced reaction while an existing one is already in place.
	
	%==================================
	% FIGURE
	%==================================
	
	It is worth noting that our model demonstrates the capability to detect external perturbations, such as those caused by significant events like the COVID-$19$ pandemic. This capacity makes our approach valuable not only for understanding historical trends but also for proactively identifying and characterizing disruptions in the adoption and detachment patterns of cultural phenomena.
	
	\begin{table}[H]
		\centering
		\caption{Fitting parameters and the root mean square (RMS) error of the residuals for the Google Trends data for keywords ``Facebook'' and ``Snapchat''}
		\begin{tabular}{|c|c|c|c|c|c|}
			\hline
			Social Media & RMS of Res. & $\sigma$ & $\beta$ & $\alpha$ & $M_{0}$ \\ \hline
			Facebook & 5.87717 & 235.071 & 0.119982 & 4.50933 & 1.35927e+06 \\ \hline
			Snapchat & 8.3044 & 44.292 & 0.096952 & 4.39675 & 0.00122246 \\ \hline
		\end{tabular}
		\label{tab:second_fit}	
	\end{table}
	
	Moreover, our model allows us to make predictions about the business cycles. The time for an innovation to peak and vanish can be readily calculated from Eq. \ref{eq:number_of_attachments_over_time}. These results are shown in Tab. \ref{tab:peakext}.
	
	\begin{table}[H]
		\centering
		\caption{Expected time to peak and extinction for the influence of scientific papers A--C and social media ``Facebook'' and ``Snapchat''}
		\begin{tabular}{|c|c|c|c|c|c|}
			\hline
			Innovation & A & B & C & `Snapchat' & `Facebook'\\
			\hline
			Peak [years] & 11 & 13 & 29 & 5 & 9\\
			\hline
			Extinction [years] & 99 & 79 & 49 & 11 & 20\\
			\hline
		\end{tabular}
		\label{tab:peakext}
	\end{table}
		
%================================================
% SECTION
%================================================	
\section{Estimating $\lambda_0$ and $\mu_0$}
A good way to test our ansatz for birth and death functions is to use a neural network to estimate them and compare the results with our theoretical predictions.

In our analytical model, the birth and death rates can be discretized to:

\begin{eqnarray}
	\lambda_0^{(m)} &=& \frac{\alpha}{m}\nonumber,\\
	\mu_0^{(m)} &=& \beta\left(1-e^{-m/\sigma}\right).
	\label{eq:bdrates}
\end{eqnarray}
With these equations, we can obtain the main trend of adoption and detachment from innovations in a way that mimics pulsed neutron-induced radioactive activation. These equations, however, only capture general trends and are unable to capture fine details of the innovation process. Moreover, they were constructed based on a set of assumptions that may not happen in real situations.

\begin{figure}[H]
	\centering
	\includegraphics[scale=0.5]{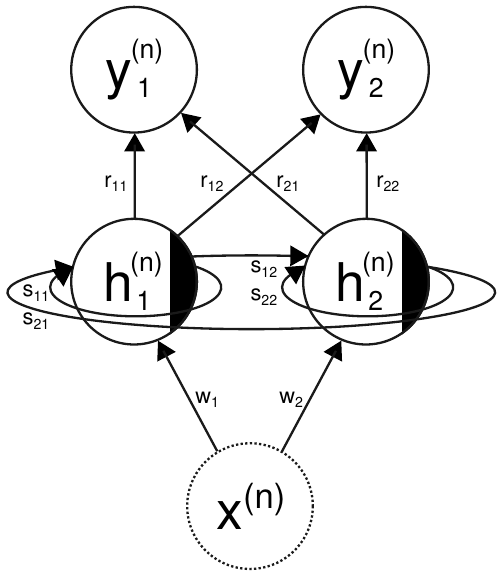}
	\caption{Recurrent neural network used to estimate $\lambda^{(n)}$ and $\mu^{(n)}$ as outputs $y_{1,2}$ given inputs $x^{n}$. The dotted circle indicates an input, the partially filled circles indicate a hyperbolic activation and the solid circle represents an identity activation.}
	\label{fig:Elman}
\end{figure}

To investigate the birth and death rates in real situations, we employ a simple Elman recurrent neural network (RNN) \cite{CRC_ML_Book,Elman} with one input $x^{(n)}$, one hidden layer $h_{1,2}^{(n)}$ with hyperbolic activations, and two linear outputs $y_{1,2}^{(n)}$, as shown in Fig. \ref{fig:Elman}. The layers are mathematically given as:

\begin{eqnarray}
	h_1^{(n)} &=& \tanh\left(w_1x+s_{11}h_1^{(n-1)}+s_{21}h_2^{(n-1)}+b_1\right)\nonumber\\
	h_2^{(n)} &=& \tanh\left(w_2x+s_{12}h_1^{(n-1)}+s_{22}h_2^{(n-1)}+b_1\right)\nonumber\\
	z_1^{(n)} &=& r_{11}h_1^{(n)}+r_{21}h_2^{(n)}+c_1\nonumber\\
	z_2^{(n)} &=& r_{12}h_1^{(n)}+r_{22}h_2^{(n)}+c_2,
\end{eqnarray}
where parameters $\mathbf{s}, \mathbf{r}, \mathbf{b}$ and $\mathbf{c}$ are obtained after training the network.

The realistic birth and death rates are estimated in a two-step approach. First, the network is pre-conditioned to predict theoretical rates in Eq. \ref{eq:bdrates} from synthetic data produced by \ref{eq:number_of_attachments_over_time}. This ensures that all simulations with real data depart from the same conditions and they reflect the theoretical expectation. Then, the network is allowed to relax by adjusting the predicted values of $\lambda_0$ and $\mu_0$ to values that satisfy Eq. \ref{eq:Adherence_Variation} with experimental values for $M$.

In the pre-conditioning stage, values for $\alpha,\beta$, and $\sigma$ are randomly selected. Then values of $M$ are obtained using Eq. \ref{eq:number_of_attachments_over_time}, and the network is trained to produce estimates for the corresponding $\lambda_0$ and $\mu_0$. For this purpose, the network is trained with the loss:

\begin{equation}
	\mathcal{L}_{pc} = \frac{1}{N}\sum_{i=1}^N\left[\left(y_1^{(i)}-\lambda_0^{(i)}\right)^2 +\left(y_2^{(i)}-\mu_0^{(i)}\right)^2\right],
\end{equation}	
where $N$ is the length of the time series.

With the network pre-conditioned with the theoretical values, we let it relax to new estimates given experimental data for $M$ and a loss function given by:

\begin{equation}
	\mathcal{L}_{ex} = \frac{1}{N}\sum_{i=1}^N \left[\left(\frac{dM^{(i)}/dt}{M^{(i)}}-y_1^{(i)}+y_2^{(i)}\right)^2+\delta\left(\left|y_1^{(i)}\right|+\left|y_2^{(i)}\right|\right)\right].
\end{equation}
The loss function is comprised of two terms. The first penalizes deviations from Eq. \ref{eq:Adherence_Variation}, while the second discourages large output values through regularization. In all simulations, we added noise to $M$ at a magnitude one order smaller than its minimum value. The derivative of $M$ was estimated using the Savitzky-Golay procedure \cite{Savitzky1964,Kouibia2012} with $15$ kernel nodes. Finally, all simulations employed $\delta=10^{-5}$ and the ADAM optimizer \cite{CRC_ML_Book} with parameters $\eta=3\times10^{-3}$, $\beta_1=0.9$, and $\beta_2=0.999$. Training was conducted until the Gaussian curvature of the loss function, computed with $500$ data points, reached the value of $\rho=10^{10}$. The Gaussian curvature for a dataset $\{x\}$ is defined as:

\begin{equation}
	\rho_x(t)=\left|\frac{\left(1+dx/dt\right)^{3/2}}{d^2x/dt^2}\right|.
\end{equation}

	%....................................................................
	% SUBSECTION
	%....................................................................
	\subsection{Results}
\begin{figure}[H]
		\centering
		\includegraphics[scale=0.75]{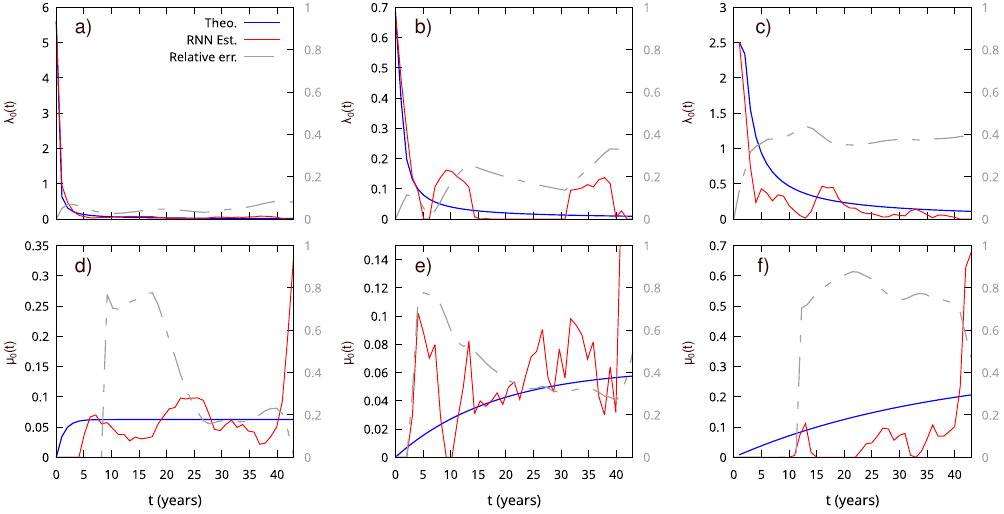}
		\caption{Estimated $\lambda_0$ and $\mu_0$ for the citations of papers A (a) and (d), B (b) and (e), and C (c) and (f). Blue solid lines correspond to the theoretical curves, while the red solid curves correspond to the estimated curves using RNN. The dashed gray lines are the relative errors.}
		\label{fig:LM1}
	\end{figure}

	Figure \ref{fig:LM1} shows the estimated birth and death rates for scientific papers A--C using both Eq. \ref{eq:bdrates} and the RNN. We quantified the relative difference between them as:
	
	\begin{equation}
		\Delta_n = \frac{\left|\sum_{i=1}^nX^{(i)}-\sum_{i=1}^ny_{1,2}^{(i)}\right|}{\sum_{i=1}^nX^{(i)}},
	\end{equation}
	where $X$ is either $\lambda_0$ or $\mu_0$.
	
	This approach of calculating the relative difference with the accumulated time series attenuates short-term fluctuations and emphasizes underlying trends through temporal integration. By mitigating the influence of transient variations, it facilitates the detection of long-term divergences and subtle shifts in trend dynamics. 
	
\begin{figure}[H]
		\centering
		\includegraphics[scale=1.2]{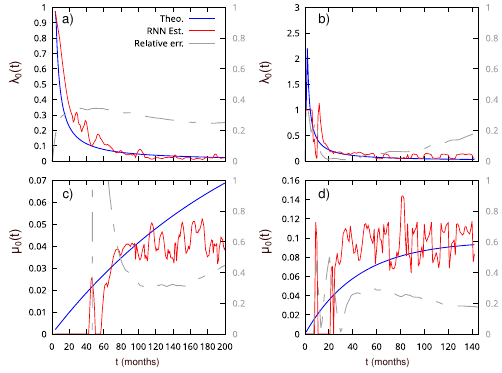}
		\caption{Estimated $\lambda_0$ and $\mu_0$ for the temporal evolution of interest in the keywords `Snapchat' (a) and (c), and `Facebook' (b) and (d). Blue solid lines correspond to the theoretical curves, while the red solid curves correspond to the estimated curves using RNN. The dashed gray lines are the relative errors.}
		\label{fig:LM2}
	\end{figure}

	The same analysis was made for the interest in keywords ``Snapchat'' and ``Facebook''. These results are shown in Fig. \ref{fig:LM2}.
	
	In a neutron-induced radioactive process, bursts may induce transitions between different decay modes or interact with various isotopes or nuclear states, resulting in shifts in the decay-active population \cite{modes}. Additionally, resonances in nuclear cross-sections can impact neutron capture, leading to oscillations in the decay rate \cite{reso1,reso2}. The decay products can exert influence on the environment, initiating feedback loops that, in turn, affect the birth and death rates \cite{feedback1,feedback2}. Finally, fission events within the system can generate additional neutrons, contributing to fluctuations in the decay rates.
	
	These phenomena find a metaphorical parallel in human behavior \cite{EuEcon}. Social interactions are susceptible to external shocks such as wars, pandemics, and crises, capable of triggering rapid behavioral changes. Career transitions and migration can further contribute to population shifts and fluctuations. Cultural trends may introduce biases that shape our perception of information \cite{social1,social2,social3}. Lastly, individual actions exhibit a feedback loop effect, resulting in complex dynamics and oscillations in behavior.
	
	Our model demonstrates the ability to capture the outcomes of these intricate phenomena and estimate their impact on birth and death rates. To investigate
	the origin of these features in more depth, we propose an automaton that mimics both the behaviors of radioactive and innovation processes.
	
%================================================
% SECTION
%================================================	
\section{Capturing the Observed Behavior with an Automaton}
\label{chap:Relationship_with_other_Models}
As discussed in the previous section, the dynamics of innovation can be a complex process often involving multiple interacting factors. In this last section, we propose an automaton that effectively captures the main trends of the observed behaviors through a set of simplified rules. This approach allows us to verify our findings, explore different scenarios, identify key elements of the system, and represent other complex systems that might have potential similarities.

Our automaton is described by a tuple $A=(Z,S,N,f)$ where $Z$ is a $d\times d\in\mathbb{N}^2$ lattice composed of discrete cells that can hold states $S=\{1,2,3\}$. The transition from one cell value to another
is given by the transition function $f:S\rightarrow S$. The transition of states for a single cell is considered within a von Neumann neighborhood, described by the set $N(\mathbf{n}_0)=\{\mathbf{n}_k:\sum_i^d|(\mathbf{n}_k)_i-(\mathbf{n}_0)_i\leq 1\}$.
The configuration of the automaton
is described by a function $c:Z\rightarrow S$ that assigns a state to each cell. The next state of a cell is given by
$c(\mathbf{n}_0)^{t+1}=f(\{\mathbf{n}_m|\mathbf{n}_m\in N(\mathbf{n}_0)^t\}$.
The next state for a cell is determined through a random selection from the values of its neighboring cells, with the selection probabilities being proportional
to the frequencies of those values.
Therefore, each agent tends to align its state to the state of the majority of its neighbors. To avoid boundary effects, we used periodic boundary conditions.

\begin{figure}[H]
	\centering
	\includegraphics{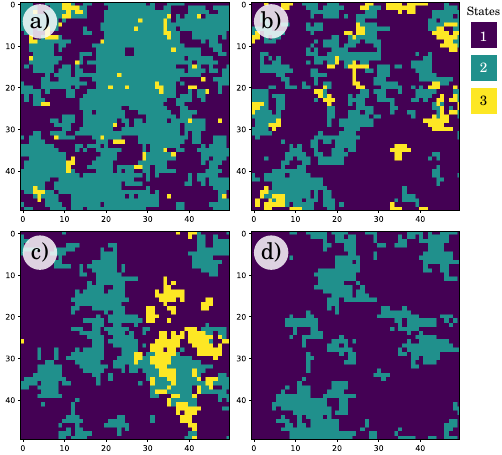}
	\caption{Output grid of the automaton: a) immediately after introducing a new state, b) after $10$ steps, c) after $30$ steps, and d) after $50$ steps.}
	\label{fig:snap}
\end{figure}

In our simulations, the grid is randomly initiated with $S_0=\{1,2\}$. The automaton runs for $100$ steps to reach thermalization and a new state is
introduced to $n$ elements of the grid to create $S=S_0\cup\{3\}$ at random locations on the lattice. Snapshots taken immediately after introducing a new state, as well as after $10$, $30$, and $50$ steps, are shown in Fig. \ref{fig:snap}.

\begin{figure}[H]
	\begin{center}
		\includegraphics[scale=1]{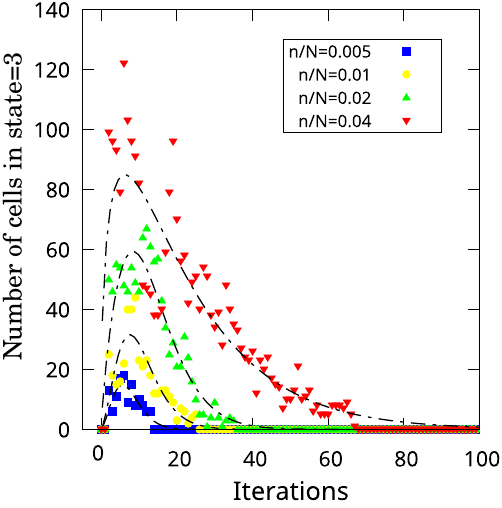}
		\caption{Number of agents in state $S=\{3\}$ as a function of Monte Carlo iteration steps on a $50\times 50$ lattice. $n/N$ represents the ratio between cells starting in state $3$ after thermalization and the total number of cells. Dashed lines represent the analytical solution of our innovation model given by a fitting of Eq. \ref{eq:number_of_attachments_over_time} with data generated by the automaton.}
		\label{fig:automata_dynamics}
	\end{center}
\end{figure}

To verify the relationship of the automata with our model, we ran a series of simulations on a $50\times 50$ lattice, and a fraction $\frac{n}{N}=\{0.005$,  $0.01$, $0.02$, $0.04\}$ of $n$ agents initialized with the state $S=\{3\}$ after the thermalization process. We then count the number of agents in state $S=\{3\}$ after each simulation step. The temporal evolution of cells in state $3$ is shown in Fig. \ref{fig:automata_dynamics}. 

Fitting the number of cells in state $S=\{3\}$ using our model, given by equation \ref{eq:number_of_attachments_over_time}, using Levenberg-Marquadt’s procedure, gives the dashed lines in Fig. \ref{fig:automata_dynamics}. The results visually match the model, and are confirmed by parameters and errors shown in Table \ref{tab:automata_fitting_results}.

\begin{table}[H]
	\centering
	\caption{Fitting parameters and the root mean square (RMS) error of the residuals for the data generated by the automaton}
	\begin{tabular}{|c|c|c|c|c|c|}
		\hline
		n/N   & RMS of Res. & $\sigma$ & $\beta$    & $\alpha$ & $M_{0}$     \\ \hline
		0.005 & 1.47763     & 3.04674  & 0.247942   & 0.975034 & 10.9288     \\ \hline
		0.01  & 3.1669      & 8.28948  & 0.314912   & 1.43169  & 53.8677     \\ \hline
		0.02  & 5.0286      & 15.1689  & 0.21077    & 0.754141 & 439.929     \\ \hline
		0.04  & 12.6278     & 1.63855  & 0.0610844  & 0.374988 & 62.7025     \\ \hline
	\end{tabular}
	\label{tab:automata_fitting_results}
\end{table}

\begin{figure}
	\centering
	\includegraphics[scale=0.85]{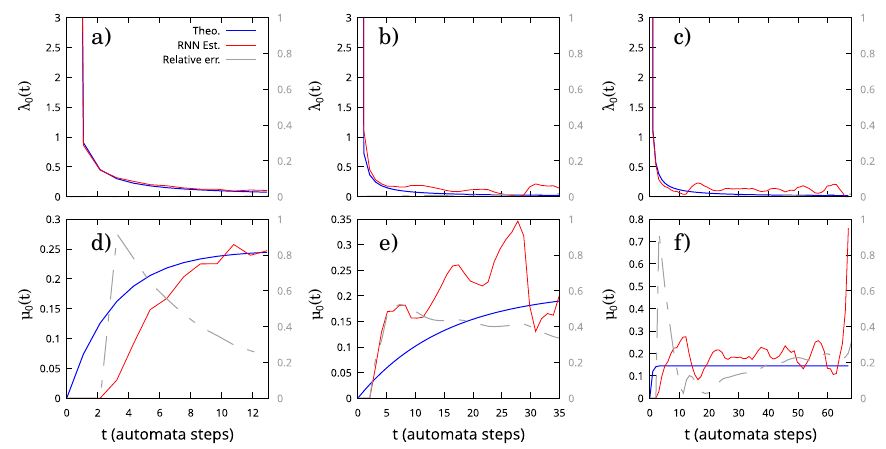}
	\caption{Estimated $\lambda_0$ and $\mu_0$ for the evolution of three-state automata with different fractions of cells initialized in a new state $\{3\}$: $0.5$ \% a) and d), $2$ \% b) and e), and $4$ \% c) and f). Blue solid curves correspond to theoretical values, while the red curves correspond to the estimated curves using RNN. The dashed gray curves are the relative errors.}
	\label{fig:autores}
\end{figure}

We used the results shown in Fig. \ref{fig:automata_dynamics} as inputs for the RNN and obtained estimates for $\mu_0(t)$ and $\lambda_0(t)$ which are shown in Fig. \ref{fig:autores}.

The fact that our automaton model can successfully capture major trends of both radioactive and innovation processes primarily suggests that they share similar underlying physics related to the birth and death of species or to the attachment and detachment of agents to new states. Also, the good fitting with long-term trends suggests that our estimates for the temporal evolution of birth and death functions are reasonable. Finally, the oscillations and deviances we see from the long-term trends are likely natural features caused by the stochastic nature of the processes involved. Although we can estimate their impact on the birth and death curves, they are the result of uncertainties related to the degrees of liberties intrinsic to the processes. Nonetheless, large deviations from the long-term trends, as depicted in Fig. \ref{fig:social_media} can indicate the application of external perturbations or the onset of new competing processes.

%================================================
% SECTION
%================================================	
\section{Conclusions}
In this work, we have shown how a Markovian innovation model can capture the behaviors of both radioactive decay processes and innovation processes. We built upon the Bass diffusion model and used simple ansatzes to create a birth-death Markov process that explains the long-term trend of innovation processes. To find the birth and death functions we used two simple ansatzes. First, we argue that the birth function must be high at the beginning of the process, but as the material is consumed it must vanish. Second, we argue that the death function must be akin to a mean-reverting process.
Given these birth and death curves obtained under these assumptions, we found a Gompertz-like function that models the number of active species.

To test our model, we applied it to two different innovation processes, namely the number of citations for three physics papers, and cultural trends captured by Google searches for trend words ``facebook'' and ``snapchat''. Our model correctly captures the long-term trends for all these cases.

To check the validity of our ansatzes for the birth and death curves, we used a recurrent neural network model that was previously trained on the theoretical model but is allowed to relax to fit real data. The neural network confirmed the long-term trends of our model and showed short-term oscillations and deviations for the birth and death curves.

The oscillations in the model were probed deeper with an automaton created to model both radioactive and innovation processes. The results obtained with the automaton show the same oscillation features observed with real data. The good fitting obtained with the automaton suggests that both processes share similar underlying physics, and the oscillations are caused by the intrinsic stochastic behavior of the processes involved.

%================================================
% SECTION
%================================================	
\bibliographystyle{unsrt}
\bibliography{innovation}

\begin{thebibliography}{10}

\bibitem{nuclear1}
M{\'a}t{\'e} Hal{\'a}sz and M{\'a}t{\'e} Szieberth.
\newblock Markov chain models of nuclear transmutation: Part i -- theory.
\newblock {\em Ann. Nucl. Energy}, 121:429--445, 2018.

\bibitem{Bass}
F.~Bass.
\newblock A new product growth for model consumer durables.
\newblock {\em Manage. Sci.}, 15(5):215--227, 1969.

\bibitem{Bass2}
J{\'e}r{\^o}me Massiani and Andreas Gohs.
\newblock The choice of bass model coefficients to forecast diffusion for
  innovative products: An empirical investigation for new automotive
  technologies.
\newblock {\em Res. Transp. Econ.}, 50:17--28, 2015.
\newblock Electric Vehicles: Modelling Demand and Market Penetration.

\bibitem{Bass3}
Hakyeon Lee, Sang~Gook Kim, Hyun woo Park, and Pilsung Kang.
\newblock Pre-launch new product demand forecasting using the bass model: A
  statistical and machine learning-based approach.
\newblock {\em Technol. Forecast. Soc. Change}, 86:49--64, 2014.

\bibitem{IdeaNet}
Iacopo Iacopini, Stasa Milojevic, and V.~Latora.
\newblock Network dynamics of innovation processes.
\newblock {\em Phys. Rev. Lett.}, 120 4:048301, 2017.

\bibitem{catas}
P.~Herbig.
\newblock A cusp catastrophe model of the adoption of an industrial innovation.
\newblock {\em J. Prod. Innov. Manage.}, 8:127--137, 1991.

\bibitem{PRE}
Shinsuke Shimogawa, Miyuki Shinno, and Hiroshi Saito.
\newblock Structure of s-shaped growth in innovation diffusion.
\newblock {\em Phys. Rev. E}, 85:056121, May 2012.

\bibitem{InoRev}
P.A Geroski.
\newblock Models of technology diffusion.
\newblock {\em Res. Policy}, 29(4):603--625, 2000.

\bibitem{InoRev2}
Mariangela Guidolin and Piero Manfredi.
\newblock Innovation diffusion processes: Concepts, models, and predictions.
\newblock {\em Annu. Rev. Stat. Appl.}, 10(1):451--473, 2023.

\bibitem{InoRev3}
Joseph~J. Jacobsen and Stephen~J. Guastello.
\newblock Diffusion models for innovation: s-curves, networks, power laws,
  catastrophes, and entropy.
\newblock {\em Nonlinear Dynamics Psychol. Life Sci.}, 15(2):307--33, 2011.

\bibitem{Gompertz}
B.~Gompertz.
\newblock On the nature of the function expressive of the law of human
  mortality, and on a new mode of determining the value of life contingencies.
\newblock {\em Phil. Trans. Roy. Soc. London}, 123:513--585, 1832.

\bibitem{Banerjee1992}
A.~V. Banerjee.
\newblock A simple model of herd behavior.
\newblock {\em Q. J. Econ.}, 107(3):797--817, 1992.

\bibitem{EuEcon}
C.~R. da~Cunha.
\newblock {\em Introduction to Econophysics, Contemporary approaches with
  python simulations}.
\newblock CRC Press, Boca Raton, FL, 2022.

\bibitem{neutron2}
Alexander~P. Barzilov, Ivan~S. Novikov, and Brian Cooper.
\newblock Computational study of pulsed neutron induced activation analysis of
  cargo.
\newblock {\em J. Radioanal. Nucl. Chem.}, 282(1):177--181, Oct 2009.

\bibitem{neutron3}
N.~Colonna, F.~Gunsing, and F.~K{\"a}ppeler.
\newblock Neutron physics with accelerators.
\newblock {\em Prog. Part. Nucl. Phys.}, 101:177--203, 2018.

\bibitem{neutron4}
Paul Kehler.
\newblock Pulsed neutron measurement of single and two-phase liquid flow.
\newblock {\em IEEE Trans. Nucl. Sci.}, 26(1):1627--1631, 1979.

\bibitem{neutron1}
W.~D. James and J.~A. Oyedele.
\newblock Application of reactor pulsing to neutron activation analysis.
\newblock {\em J. Radioanal. Nucl. Chem.}, 110(1):33--40, Mar 1987.

\bibitem{cspace}
M.~H. Zwietering, I.~Jongenburger, F.~M. Rombouts, and K.~van't Riet.
\newblock Modeling of the bacterial growth curve.
\newblock {\em Appl. Environ. Microbiol.}, 56(6):1875--1881, 1990.

\bibitem{CRC_ML_Book}
C.~R. da~Cunha.
\newblock {\em Machine Learning for the Physical Sciences: Fundamentals and
  Prototyping with Julia}.
\newblock CRC Press, Boca Raton, FL, 2023.

\bibitem{LeGuillou1980}
J.~C.~Le Guillou and J.~Zinn-Justin.
\newblock Critical exponents from field theory.
\newblock {\em Phys. Rev. B}, 21:3976, 5 1980.

\bibitem{Ambegaokar1980}
Vinay Ambegaokar, B.~I. Halperin, David~R. Nelson, and Eric~D. Siggia.
\newblock Dynamics of superfluid films.
\newblock {\em Phys. Rev. B}, 21:1806, 3 1980.

\bibitem{Klein1980}
B.~M. Klein, D.~A. Papaconstantopoulos, and L.~L. Boyer.
\newblock
  Linear-combination-of-atomic-orbitals-coherent-potential-approximation
  studies of carbon vacancies in the substoichiometric refractory monocarbides.
\newblock {\em Phys. Rev. B}, 22:1946, 8 1980.

\bibitem{Lee2022}
E.~D. Lee, C.~P. Kempes, and G.~B. West.
\newblock Idea engines: A unified theory of innovation and obsolescence from
  markets and genetic evolution to science.
\newblock {\em Proc. Natl. Acad. Sci.}, 121(6):e2312468120, 2024.

\bibitem{Elman}
Jeffrey~L. Elman.
\newblock Finding structure in time.
\newblock {\em Cogn. Sci.}, 14(2):179--211, 1990.

\bibitem{Savitzky1964}
Abraham Savitzky and Marcel~J.E. Golay.
\newblock Smoothing and differentiation of data by simplified least squares
  procedures.
\newblock {\em Anal. Chem.}, 36:1627--1639, 7 1964.

\bibitem{Kouibia2012}
Abdelouahed Kouibia and Miguel Pasadas.
\newblock An approximation problem of noisy data by cubic and bicubic splines.
\newblock {\em Appl. Math. Model.}, 36:4135--4145, 9 2012.

\bibitem{modes}
Bernd Crasemann.
\newblock Some aspects of atomic effects in nuclear transitions.
\newblock {\em Nucl. Instrum. Methods}, 112(1):33--39, 1973.

\bibitem{reso1}
S.~Goriely.
\newblock Radiative neutron captures by neutron-rich nuclei and the r-process
  nucleosynthesis.
\newblock {\em Phys. Lett. B}, 436(1):10--18, 1998.

\bibitem{reso2}
W.~Potzel, U.~van B{\"u}rck, P.~Schindelmann, G.~M. Kalvius, G.~V. Smirnov,
  E.~Gerdau, Yu.~V. Shvyd'ko, H.~D. R{\"u}ter, and O.~Leupold.
\newblock Investigation of radiative coupling and of enlarged decay rates of
  nuclear oscillators.
\newblock {\em Phys. Rev. A}, 63:043810, Mar 2001.

\bibitem{feedback1}
Richard Cornez.
\newblock Birth and death processes in random environments with feedback.
\newblock {\em J. Appl. Probab.}, 24(1):25--34, 1987.

\bibitem{feedback2}
Boris~L. Granovsky and Alexander~I. Zeifman.
\newblock The decay function of nonhomogeneous birth-death processes, with
  application to mean-field models.
\newblock {\em Stoch. Process. Their Appl.}, 72:105--120, 1997.

\bibitem{social1}
Matthew~A. Baum and Phil Gussin.
\newblock In the eye of the beholder: How information shortcuts shape
  individual perceptions of bias in the media.
\newblock {\em Quart. J. Polit. Sci.}, 3(1):1--31, 2008.

\bibitem{social2}
Alex Mesoudi, Andrew Whiten, and Robin Dunbar.
\newblock A bias for social information in human cultural transmission.
\newblock {\em Br. J. Psychol.}, 97(3):405--423, 2006.

\bibitem{social3}
M.~B.~Fallin Hunzaker.
\newblock Cultural sentiments and schema-consistency bias in information
  transmission.
\newblock {\em Am. Sociol. Rev.}, 81(6):1223--1250, 2016.

\end{thebibliography}

\end{document}